\documentclass[preprint]{revtex4}%
\usepackage{amsfonts}
\usepackage{amsmath}
\usepackage{amssymb}
\usepackage[dvips]{graphicx}
\usepackage{graphicx}
\usepackage{float}
\usepackage{epsfig}
\usepackage{subfigure}%
\begin{document}
\preprint{ }
\title{Prediction of topological Nernst effect in silicene and similar 2D materials }
\author{I. Ahmed$^{1}$, M. Tahir$^{2}$, K. Sabeeh$^{1\ast}$}
\affiliation{$^{1}$Department of Physics, Quaid-i-Azam University, Islamabad 45320, Pakistan}
\affiliation{$^{2}$Department of Physics, University of Sargodha, Pakistan}
\affiliation{}
\affiliation{}
\affiliation{}

\begin{abstract}
We consider Berry phase mediated Nernst effect in silicene. The low energy
band structure of silicene consists of two valleys near the Dirac points,
similar to graphene. The low energy transport properties of the quasiparticles
can be described as Berry phase dependent phenomena. By contrast to graphene,
silicene has strong spin-orbit interaction leading to opening of the gap in
the energy spectrum and spin-splitting of the bands in each valley. If an
electric field is applied perpendicular to the silicene sheet, it allows
tunability of the gap. \ We show that this results in Berry-phase-supported
spin and valley polarized Nernst effect when the system is subjected to a
temperature gradient. The Nernst response can be used to create valley and
spin polarization at the transverse edges of silicene sheet. The applied
electric field also allows control of valley and spin polarization in
silicene. The predicted valley and spin polarized Nernst effect in silicene is
more general and applies to other two-dimensional (2D) buckled Dirac Fermion
systems such as 2D germanium and tin.

Pacs:

\end{abstract}
\startpage{01}
\endpage{02}
\maketitle

\section{Introduction}

Silicene is a monolayer of silicon atoms\cite{1}, forming a buckled honeycomb
structure \cite{2,3}. It does not occur naturally but it has been synthesized
on metal surfaces\cite{4,5}. Its low energy band structure is similar to
graphene with two inequivalent valleys at the Dirac points. By contrast to
graphene, it has large Spin-Orbit Interaction (SOI) being more pronounced in
silicon than in carbon\cite{3}, which induces a gap of 3.9meV, for graphene
this gap is around 24$\mu$eV\cite{6}. This significant SOI gives rise to two
important phenomena. First, by opening the gap, it provides mass to the Dirac
fermions\cite{7,8}. Second, it spin-splits the bands in each
valley\cite{7,8,9}. The latter property makes possible the manipulation of the
spin degree of freedom, in addition to the valley degree of freedom, with
possible applications in spintronics. Further, the gap in silicene is tunable
by applying an external uniform electric field $E_{z}$ perpendicular to the
silicene sheet\cite{7,8,9,10,11}. Interestingly, it has been shown that
silicene shows a number of topologically protected phases, when subjected to
the field $E_{z}$\cite{7,8,11}. More importantly for this work, $E_{z}$-field
breaks the Space Inversion (SI) symmetry of the system. Hence the Berry
curvature is finite and sharply peaked at the two valleys. This gives rise to
nontrivial topological electric transport phenomena. In graphene, it led to
the prediction of valley Hall effect \cite{12}.

In this paper, we investigate the Berry phase effects on another class of
transport coefficients, the thermoelectric transport coefficients. Our main
focus is on topological Nernst effect. This Berry phase mediated Nernst effect
exists independent of an external magnetic field. Prior to our work and
relevant to it, there has been investigation of Nernst effect in gapped single
and bilayer graphene\cite{13} based on Berry phase formulation developed
earlier\cite{14,15}. In both gapped graphene systems, the Nernst effect is
valley dependent and can lead to valley polarization. By contrast in silicene,
we will show that Nernst effect is valley as well as spin dependent.
Therefore, in a generic Nernst measurement set up it can be used to generate
valley as well as spin polarization which can have important technological implications.

The main question that we address, in this work, is the valley and spin
tunability of the Nernst effect by an applied electric field. A related
question is the ability to generate valley and spin polarization in silicene.
These questions are important because silicene has rich physics, a variety of
different phases\cite{7,8}, and compatability with present silicon
microelectronics. Further, it has been recently shown that that the figure of
merit which quantifies the thermoelectric effeciency of a material is much
higher for silicene compared to graphene\cite{23}. Therefore, it is believed,
that silicene can be a better option for electrically tunable thermoelectric
devices than graphene\cite{10,11}.

\section{Berry Curvature and Magnetic Moment}

Silicene has a honeycomb structure of silicon atoms on sublattice sites A and
B displaced from each other by a distance 2$\mathit{l.}$ The graphene-like low
energy effective Hamiltonian of silicene, in the presence of SOI and subjected
to a perpendicular $E_{z}$-field around Dirac point within each valley, can be
expressed around the $K_{\eta}$ point as\cite{7,8}%

\begin{equation}
H_{\eta}^{s_{z}}=\hbar\upsilon\left(  -\eta q_{x}\sigma_{x}+q_{y}\sigma
_{y}\right)  +\eta s_{z}\Delta_{so}\sigma_{z}+\Delta_{z}\sigma_{z}%
\end{equation}
where, $\eta=+/-$ for $K_{+,-}$ Dirac points (valley index), $\Delta
_{z}=\mathit{l}E_{z}$, with \textit{l}=0.23\.{A}, $\left(  \sigma_{x}%
,\sigma_{y},\sigma_{z}\right)  $ are Pauli matrices, $\upsilon$\ is the Fermi
velocity of Dirac fermions and ($q_{x},q_{y})$ are the components of the wave
vector relative to the Dirac point. The spin index $s_{z}=+/-$ for spin up
$\left(  \uparrow\right)  $\ and down $\left(  \downarrow\right)
,$\ respectively. In the absence of any $\widehat{s}_{z}$ nonconserving terms
in the Hamiltonian, one can simply consider the two $2\times2$ subspaces
corresponding to $s_{z}=\pm1$ separately. The first term in the Hamiltonian,
arises from nearest neighbour hopping and is the Dirac term, well known from
studies in graphene. The second term is the intrinsic spin-orbit interaction
term (Kane and Mele term) with spin-orbit gap of $\Delta_{so}=3.9meV.$ The
third term is due to the uniform perpendicular electric field $E_{z}$ with the
electric field induced gap $\Delta_{z}.$ It generates a staggered sublattice
potential between the sites A and B.

The Hamiltonian can be diagonalized analytically, and its energy spectrum is%
\begin{equation}
\varepsilon_{\eta}^{s_{z}}=\pm\sqrt{\hbar^{2}\upsilon^{2}q^{2}+\left(
\Delta_{z}+\eta s_{z}\Delta_{so}\right)  ^{2}}%
\end{equation}
where +(-) solution is for electron(hole) bands. As we mentioned earlier, our
model Hamiltonian Eq.(1) contains an external tunable parameter $\Delta_{z}$,
which can be tuned relative to $\Delta_{so}$. So we have three possible
situations: $\Delta_{z}$ is less than $\Delta_{so}$, $\Delta_{z}$\ is equal to
$\Delta_{so}$\ and $\Delta_{z}$\ is greater than $\Delta_{so}$. The energy
dispersion, Eq. (2), for the three cases is presented in Fig.(1a), Fig.(1b)
and Fig.(1c).

The eigenvectors for electron and hole states in $K_{+}$valley are $\left\vert
\uparrow_{K+}^{e}\right\rangle =%
\begin{pmatrix}
-\cos\frac{\theta}{2}e^{\mathit{i}\phi}\\
\sin\frac{\theta}{2}%
\end{pmatrix}
\exp[iq_{x}x+iq_{y}y]$ and $\left\vert \uparrow_{K_{+}}^{h}\right\rangle =%
\begin{pmatrix}
\sin\frac{\theta}{2}e^{\mathit{i}\phi}\\
\cos\frac{\theta}{2}%
\end{pmatrix}
\exp[iq_{x}x+iq_{y}y].$ $\theta=\tan^{-1}\frac{\hbar\upsilon q}{\Delta
_{z}+\eta s_{z}\Delta_{so}}$ and $\varphi=\tan^{-1}\frac{q_{y}}{q_{x}}$ here.
The states in $K_{-}$ valley are time-reversed conjugates of the above
states$.$

At this stage, it is necessary to discuss symmetry of our system because it
dictates whether Berry curvature is finite and nonzero or not. This is
important because topological transport will survive only when Berry curvature
does not vanish at all points in the momentum space. Berry curvature is an odd
function of $q$ in the presence of Time-Reversal (TR) symmetry and an even
function in the presence of Space Inversion (SI) symmetry\cite{15,16}. If both
these symmetries are present, Berry curvature vanishes at all points in
momentum space. The $\Delta_{z}$ term explicitely breaks SI symmetry, see
Eq.(1)\cite{12,15,16}. The TR symmetry when both the Dirac points is
considered is intact. In this case, Berry curvature is finite and nontrivial
which will have profound consequences for transport in the system. From the
Hamiltonian, Eq. (1), and the eigenvectors, the Berry curvature for the
conduction band (electrons), can be found:%

\begin{equation}
\Omega_{\eta}^{s_{z}}=\left(  \eta\frac{\hbar^{2}\upsilon^{2}}{2}\right)
\frac{\left(  \Delta_{z}+\eta s_{z}\Delta_{so}\right)  }{\left(  \hbar
^{2}\upsilon^{2}q^{2}+\left(  \Delta_{z}+\eta s_{z}\Delta_{so}\right)
^{2}\right)  ^{\frac{3}{2}}}.
\end{equation}
Note that Berry curvature has opposite signs in the two valleys for opposite
spins as required by TR symmetry: $\Omega_{+}^{+}=-\Omega_{-}^{-}$ and
$\Omega_{-}^{+}=-\Omega_{+}^{-}.$

In an important study on valley contrasting physics in graphene\cite{12}, it
was found that one of the properties that distinguishes valley degree of
freedom is the magnetic moment. It is of orbital nature since SOI is weak in
graphene. It depends on the valley index and can be used to create valley
polarization in the presence of an external magnetic field. By contrast to
graphene, SOI is relatively large in silicene. It spin splits the bands in
silicene\cite{7,8,9}. Therefore, in silicene, the Bloch fermions carry the
spin magnetic moment in addition to the orbital magnetic moment which
originates from the self rotation of their wave packet. It is important to
note that valley magnetic moment in silicene has both valley and spin
character. The orbital magnetic moment can be obtained from a semi-classical
formulation of wave-packet dynamics and is given by\cite{15,17,18}
\begin{equation}
\overrightarrow{m}\left(  \overrightarrow{q}\right)  =-\mathit{i}\left(
\frac{e}{2\hbar}\right)  \left\langle \overrightarrow{\nabla}_{q}%
u(q)\right\vert \times\left[  H(q)-\varepsilon(q)\right]  \left\vert
\overrightarrow{\nabla}_{q}u(q)\right\rangle
\end{equation}
where $\left\vert u(q)\right\rangle $\ is the periodic part of the Bloch
function, $H(q)$\ is the Bloch Hamiltonian and $\varepsilon(q)$\ is band
dispersion. In analogy with graphene, since it depends on the valley index, it
can be called the Valley Magnetic Moment (VMM). But we must remember that in
silicene, unlike graphene, it is also a spin dependent quantity. For a 2D
sheet the VMM is always normal to the plane. It comes out to be%
\begin{equation}
m_{\eta}^{s_{z}}=\eta\left(  \frac{e}{\hbar}\right)  \left(  \frac{\hbar
^{2}\upsilon^{2}}{2}\right)  \frac{\left(  \Delta_{z}+\eta s_{z}\Delta
_{so}\right)  }{\left(  \hbar^{2}\upsilon^{2}q^{2}+\left(  \Delta_{z}+\eta
s_{z}\Delta_{so}\right)  ^{2}\right)  }.
\end{equation}
It is concentrated at the Dirac points in the valleys. Furthermore, the VMM of
the two valleys depends on the valley index. At the bottom of the band, where
$q\rightarrow0$, and for the parameters $\Delta_{z}=\Delta_{so}\sim4$meV and
$\upsilon\sim5.5\times10^{5}$m/s,\ the VMM is about twice that of graphene (it
is 63.78$\mu_{B}$ for silicene whereas for graphene it is 30$\mu_{B}$). This
means, by contrast to graphene, we expect a stronger response in silicene to
an applied perpendicular magnetic field. An applied magnetic field will couple
to the valley magnetic moment. A net valley polarization, more moments in one
valley compared to the other (population difference), will be achieved. The
result is that an enhanced Pauli paramagnetism like phenomena with larger net
magnetization is expected compared to graphene\cite{12,16}. Therefore,
silicene is a better option to realize valley as well as spin polarization
than graphene.

\section{Charge Valley and Spin Hall Conductivity}

To determine the Nernst effect, we will proceed in two directions. First, we
will determine the charge Hall conductivity and using the Mott relation arrive
at the Nernst conductivity $\alpha_{xy}$. This has the advantage that one can
find an analytic expression for $\alpha_{xy}$. The drawback is that the result
is restricted to low temperature. The second approach is to employ the Berry
formalism to compute the Nernst conductivity using the entropy density of the
system. In this approach, one usually can not obtain analytic results at
finite temperature because of the Fermi function integrals. At the end
numerical integration is required.

\textbf{Valley Hall Conductivity:} We will begin with the first approach that
requires calculating the charge Hall conductivity. This we can do by employing
the Berry phase formalism. We are ignoring impurity scattering here and
consider only the intrinsic contribution. In the presence of an in-plane
electric field, fermions acquire anomalous velocity proportional to the Berry
curvature that gives rise to the intrinsic Hall conductivity\cite{18,19,20}.
For a particular valley (say $K_{+})$ and band $n$, the Valley Hall
Conductivity (VHC) is given by
\begin{equation}
\sigma_{xy}^{v,n}=\frac{e^{2}}{\hbar}\int_{0}^{q_{F}}\frac{d^{2}q}{\left(
2\pi\right)  ^{2}}\left[  f_{\eta=+1,n}^{s_{z}=+1}(\varepsilon_{q}%
)\Omega_{\eta=+1}^{s_{z}=+1}(q)+f_{\eta=+1,n}^{s_{z}=-1}(\varepsilon
_{q})\Omega_{\eta=+1}^{s_{z}=-1}(q)\right]
\end{equation}
where $f_{\eta,n}^{s_{z}}(\varepsilon_{q})$\ is the Fermi-Dirac distribution
function for band $n$, spin $s_{z}$ in valley $\eta$. In silicene, this
depends on the valley index $\eta$, where $\eta=+1$ in the above equation, as
well as the spin index $s_{z}.$ The total valley Hall conductivity is the sum
over all occupied bands for both valleys. We have determined the VHC,
$\sigma_{xy}^{v},$ when the chemical potential $\mu$ is placed in the band
gap, between the two spin split conduction bands (one band partially occupied)
and above the bottom of both the conduction bands (both bands are partially
occupied) for the following two cases: $\Delta_{z}$
$<$
$\Delta_{so}$ and $\Delta_{z}$
$>$
$\Delta_{so}.$ The results, obtained from Eq.(6), are shown in the following
table (1):%
\[%
\begin{tabular}
[c]{|c|c|c|}\hline
$\mu$ & $\Delta_{z}$
$<$
$\Delta_{so}$ & $\Delta_{z}$
$>$
$\Delta_{so}$\\\hline
in the band gap & $0$ & $-\frac{e^{2}}{h}$\\\hline
between spin split bands & $-\frac{e^{2}}{2h}\left[  1-\frac{\left(
\Delta_{so}-\Delta_{z}\right)  }{\mu}\right]  $ & $-\frac{e^{2}}{2h}\left[
1-\frac{\left(  \Delta_{so}-\Delta_{z}\right)  }{\mu}\right]  $\\\hline
above the bottom of both bands & $-\frac{e^{2}}{2h}\left(  \frac{2\Delta_{z}%
}{\mu}\right)  $ & $-\frac{e^{2}}{2h}\left(  \frac{2\Delta_{z}}{\mu}\right)
$\\\hline
\end{tabular}
\ \ \ \
\]
These results highlight two important points. First, $\sigma_{xy}^{v}$\ for
completely occupied bands has its maximum value $\frac{e^{2}}{2h}$, for
partially filled bands it is unquantized. Second, it increases with $E_{z}%
$-field (or $\Delta_{z}$) which allows tunability of $\sigma_{xy}^{v}$. We
also need to discuss the situation when $\Delta_{z}$ = $\Delta_{so}$. In this
case, the system is gapless, the gap closes for spin-down bands, as shown in
Fig. (1b). Eq. (3) suggests that for these bands, the Berry curvature vanishes
and they do not contribute to $\sigma_{xy}^{v}$. The contribution comes from
the spin-up bands. For a completely filled band $\sigma_{xy}^{v}$ is
($\frac{e^{2}}{2h}$).

\textbf{Spin Hall Conductivity:} The Spin Hall Conductivity (SHC) for a single
valley and band is%

\begin{equation}
\sigma_{xy}^{s,n}=\frac{e^{2}}{\hbar}\int_{0}^{q_{F}}\frac{d^{2}q}{\left(
2\pi\right)  ^{2}}\left[  f_{\eta=+1,n}^{s_{z}=+1}(\varepsilon_{q}%
)\Omega_{\eta=+1}^{s_{z}=+1}(q)-f_{\eta=+1,n}^{s_{z}=-1}(\varepsilon
_{q})\Omega_{\eta=+1}^{s_{z}=-1}(q)\right]
\end{equation}

The results, obtained from Eq.(7), are shown in the following table (2):%
\[%
\begin{tabular}
[c]{|c|c|c|}\hline
$\mu$ & $\Delta_{z}$
$<$
$\Delta_{so}$ & $\Delta_{z}$
$>$
$\Delta_{so}$\\\hline
band gap & $-\frac{e^{2}}{h}$ & $0$\\\hline
between spin splitted bands & $-\frac{e^{2}}{2h}\left[  1+\frac{\left(
\Delta_{so}-\Delta_{z}\right)  }{\mu}\right]  $ & $-\frac{e^{2}}{2h}\left[
1+\frac{\left(  \Delta_{so}-\Delta_{z}\right)  }{\mu}\right]  $\\\hline
above the bottom of both bands & $-\frac{e^{2}}{2h}\left(  \frac{2\Delta_{so}%
}{\mu}\right)  $ & $-\frac{e^{2}}{2h}\left(  \frac{2\Delta_{so}}{\mu}\right)
$\\\hline
\end{tabular}
\ \ \
\]
$\sigma_{xy}^{s}$\ has its maximum value $\frac{e^{2}}{2h}$ for completely
occupied bands and for partially occupied bands it is unquantized. Contrary to
$\sigma_{xy}^{v}$, it decreases with increasing external $E_{z}$-field (or
$\Delta_{z}$). For $\Delta_{z}$ = $\Delta_{so},$ similar to $\sigma_{xy}^{v}$,
gapless bands do not contribute to $\sigma_{xy}^{s}$, where as a completely
filled band contributes $\frac{e^{2}}{2h}$.

Hence, we find that Hall conductivities, $\sigma_{xy}^{v}$ and $\sigma
_{xy}^{s},$ are valley and spin dependent phenomena in silicene. Berry
curvature has opposite sign in opposite valleys. An in-plane electric field
will not only result in accumulation of charge from opposite valleys at
opposite edges of the sample but also opposite spin.

\section{Valley and Spin Nernst Effects}

The quasiparticles in silicene carry energy in addition to charge. If silicene
is subjected to a temperature gradient, current will flow transverse to the
applied temperature gradient. This will happen even in the absence of an
applied magnetic field. In this section, we calculate this spontaneous, Berry
phase supported Nernst effect\cite{14,21,22,24}. We find that, in silicene,
the Nernst effect is a valley and spin dependent phenomena. The expression for
the current density is%
\begin{equation}
j_{x}=\alpha_{xy}\left(  -\bigtriangledown_{y}T\right)
\end{equation}
where $\alpha_{xy}$ is the Nernst conductivity\cite{14,24}. It has been shown
that $\alpha_{xy}$ is related to zero-temperature Hall conductivity through
the Mott relation \cite{14,24}:%

\begin{equation}
\alpha_{xy}(\varepsilon)=-\frac{1}{e}\int\limits_{0}^{\infty}d\varepsilon
\frac{\partial f}{\partial\varepsilon}\sigma_{xy}(\varepsilon)\frac
{\varepsilon-\mu}{T}%
\end{equation}
where $f$\ is the Fermi-Dirac distribution function, $\mu$\ is the chemical
potential at zero-temperature (Fermi energy). $\alpha_{xy}$ comes out to be%
\begin{equation}
\alpha_{xy}\simeq\frac{\pi^{2}k_{B}^{2}T}{3e}\left(  \frac{d\sigma_{xy}(\mu
)}{d\mu}\right)  .
\end{equation}
As mentioned earler, in the Hamiltonian, Eq. (1), there is an external tunable
parameter $\Delta_{z}$. It has a direct impact on zero-temperature Hall
conductivities, which in turn affect $\alpha_{xy}$.

At higher temperatures, we will need to employ the second approach where it
will be more convenient to calculate the coefficient $\overline{\alpha}_{xy}$
which determines the transverse heat current $J^{h}$ in response to the
electric field $E:J_{x}^{h}=\overline{\alpha}_{xy}E_{y}$. This is related to
the Onsager relation $\overline{\alpha}_{xy}=T\alpha_{xy}$\cite{14,19}. For a
finite Berry curvature, the quasiparticles acquire an anomalous contribution
and the coefficient of the transverse heat current. The valley Nernst
conductivity is%

\begin{equation}
\overline{\alpha}_{xy}^{v}=T\alpha_{xy}=\frac{e}{\beta\hbar}%
{\displaystyle\sum\limits_{\eta,s_{z},n}}
{\displaystyle\int}
\frac{d^{2}q}{\left(  2\pi\right)  ^{2}}\Omega_{n,\eta}^{s_{z}}(q)s_{n,\eta
}^{s_{z}}(q)
\end{equation}
and the spin Nernst conductivity is%
\begin{equation}
\overline{\alpha}_{xy}^{s}=T\alpha_{xy}=\frac{e}{\beta\hbar}%
{\displaystyle\sum\limits_{\eta,s_{z},n}}
{\displaystyle\int}
s_{z}\frac{d^{2}q}{\left(  2\pi\right)  ^{2}}\Omega_{n,\eta}^{s_{z}%
}(q)S_{n,\eta}^{s_{z}}(q)
\end{equation}
where $S_{n,\eta}^{s_{z}}(q)=-f_{n,\eta}^{s_{z}}(q)\ln f_{n,\eta}^{s_{z}%
}(q)-(1-f_{n,\eta}^{s_{z}}(q))\ln(1-f_{n,\eta}^{s_{z}}(q))$ is the entropy
density of the Dirac fermion gas. $\Omega_{n,\eta}^{s_{z}}(q)$ and $f_{n,\eta
}^{s_{z}}(q)$ are the Berry curvature and Fermi-Dirac distribution functions
for Dirac fermions in valley $\eta$ with spin $s_{z}=\pm1\ $in band $n$,
respectively. For finite temperature, these expressions will have to be
evaluated numerically.

\textbf{Valley Nernst Effect:} For a single valley we evaluated Eq. (11)
numerically,the results are plotted in Fig.(2) and Fig.(3) at T=3K and 200K,
respectively. First, we focus on low temperature results. In Fig.(2), Valley
Nernst Conductivity (VNC) $\overline{\alpha}_{xy}^{v}$\ is plotted versus
chemical potential $\mu$. As discussed earlier for completely occupied bands
$\sigma_{xy}$ is quantized in units of $\frac{e^{2}}{h}$ and for partially
occupied bands it is unquantized. From Eq. (10), we find that completely
filled bands do not contribute to $\overline{\alpha}_{xy}^{v}$, only partially
filled bands contribute. In Figs.(2a) and (2c), there are two peaks. There is
a single peak in Fig. (1b). Each peak corresponds to $\mu$ at the bottom of a
partially filled band. In Figs.(2a) and (2c), first peak corresponds to lower
band (spin-down band), where as the second peak corresponds to upper band
(spin-up band). Whether the peak is positive or negative depends on the sign
of the Berry curvature. First peak in Fig. (2a) is negative because the Berry
curvature is negative for the contributing band . In Fig.(2b), we have a
single peak, because only a single partially occupied band contributes to
$\overline{\alpha}_{xy}^{v}$.

The low temperature results can be analyzed in light of Eq. (10) which is the
central equation in the low temperature regime. $\alpha_{xy}$\ is linear in
temperature. It is proportional to the derivative of zero-temperature Hall
conductivity $\sigma_{xy}$\ with respect to $\mu$. In the band gap,
$\sigma_{xy}$ reaches its maximum value and it decreases on either side of
band gap (as $1/\mu$)\cite{15}. From table (1), it has opposite sign in
conduction and valence bands. Therefore, $\alpha_{xy}$ is discontinuous and
with a peak as $\mu$\ touches the bottom of a band and Eq. (10) shows that it
decreases as $1/\mu^{2}$\cite{13}. The magnitude of each peak depends on the
Berry curvature of respective band.

In Fig. (3), $\overline{\alpha}_{xy}^{v}$ is plotted in the high temperature
regime. In this regime, peaks in $\overline{\alpha}_{xy}^{v}$ are completely
lost. This is due to the large contribution from thermal excitations. This
occurs when $k_{B}T\symbol{126}$ $\Delta_{so}$.

\textbf{Spin Nernst Effect:} In this part we are going to discuss behaviour of
spin Nernst conductivity $\overline{\alpha}_{xy}^{s}$. We have evaluated Eq.
(12) numerically and the results are plotted in Figs.(4a,4b,4c) at T=3K, our
focus is on the low temperature regime. Spin Nernst conductivity has been
evaluated as a function of the chemical potential as $\Delta_{z}$ is varied
relative to $\Delta_{so}$. Each partially occupied band contributes a peak to
$\overline{\alpha}_{xy}^{s}$ whose magnitude and direction depends on the
magnitude and direction of the respective band's Berry curvature. Its behavior
follows that of $\sigma_{xy}^{v}$ and $\alpha_{xy}$: $\overline{\alpha}%
_{xy}^{s}$ is proportional to $1/\mu$, seen in table 2, and Eq: (10) shows
that $\overline{\alpha}_{xy}^{s}$ is proportional to $1/\mu^{2}$. While
discussing the behaviour of $\overline{\alpha}_{xy}^{s}$, it is important to
discuss a situation where both spin-up and spin-down Nernst conductivities
overlap. For a particular $\mu$, in each case, either spin-up or spin-down
contribution dominates. Hence, we can obtain finite spin-polarized
conductivity by tuning the two relevant parameters: chemical potential $\mu$
and $\Delta_{z}$ through the applied electric field.

\textbf{Experimental Realization: }Magnetothermoelectric measurement
techniques for 2D systems are well established and have been successfully
employed to investigate the intriguing properties of Dirac Fermion systems
like graphene\cite{25,26,27}. Furthermore, Nernst effect has also been studied
in systems with magnetic order such as ferromagnetic
semiconductors\cite{21,28}. Along the same lines, it is quite feasible to
carry out Nernst effect studies in silicene proposed in this work.

The valley and spin-polarized Nernst effect predicted for silicene is more
general and applies to similar low buckled 2D Dirac Fermion systems of group
IVA elements such as germanium (germanene) and tin\cite{3,10,11,29}. In this
regard, we note that the Hamiltonian in Eq.(1) can also be used to describe
germanene, which is a honeycomb structure of germanium. Here, the SOI is even
stronger (43 meV) with $l=0.33\mathring{A}$ and the analysis presented for
silicene is also valid for germanene.

\section{Summary}

We have investigated Nernst Hall conductivity in silicene in the presence of a
uniform electric field perpendicular to the silicene sheet. The electric field
allows tunability of the band gap in silicene. Due to the relatively strong
spin-orbit interaction, Nernst Hall conductivity is both valley and
spin-dependent. This is in contrast to graphene. We have employed Berry phase
formalism to detremine the Nernst effect. We show that it is possible to
generate spin as well as valley polarization in silicene. By varying the
electric field strenth relative to the spin-orbit interaction strength we can
control valley and spin polarization in silicene.

\section{Acknowledgement}

K. Sabeeh would like to acknowledge the support of the Abdus Salam
International Center for Theoretical Physics (ICTP) in Trieste, Italy through
the Associate scheme where a part of this work was completed. I. Ahmed and K.
Sabeeh further acknowledge the support of Higher Education Commission (HEC) of
Pakistan through project No. 20-1484/R\&D/09. The authors gratefully
acknowledge enlightening discussions with Sumanta Tewari, Markus Mueller and
Di Xiao during the completion of this work.

$^{\ast}$Corresponding Author: ksabeeh@qau.edu.pk

\bigskip
\end{document}